\documentclass[prl,aps,showpacs,twocolumn,unsortedaddress]{revtex4-1}
\usepackage{graphicx,bm}
\usepackage{amssymb}
\usepackage{amsmath}
\usepackage{epsfig}
\usepackage{epsf}
\usepackage[usenames]{color}
\usepackage{hyperref}

\begin{document}

\title{Domain-wall-controlled transverse spin injection}

\author{E. van der Bijl}
\email[Electronic address: ]{e.vanderbijl@uu.nl}
\author{R.E. Troncoso}
\author{R.A. Duine}
\affiliation{Institute for Theoretical Physics, Utrecht
University, Leuvenlaan 4, 3584 CE Utrecht, The Netherlands}

\date{\today}
\begin{abstract}
We propose an effect whereby an electric current along the interface between a ferromagnetic and normal metal leads to injection of pure spin current into the normal metal, if the magnetization-direction in the ferromagnet varies along the direction of current. For the specific example of a ferromagnetic domain wall, we compute the inverse spin-Hall effect voltage this spin current gives rise to when injected into a Pt layer. Furthermore, we show that this pure spin current leads to modification of the parameters that govern spin transfer and current-driven domain-wall motion, which can be use to optimize the latter in layered magnetic systems. This effect in principle enables control over the location of spin-current injection in devices.
\end{abstract}
\maketitle

\def\bx{{\bm x}}
\def\bX{{\bm X}}
\def\bk{{\bm k}}
\def\bK{{\bm K}}
\def\bq{{\bm q}}
\def\bv{{\bm v}}
\def\bxi{{\bm \xi}}
\def\bma{{\bm m}}
\def\bmu{{\bm \mu}}
\def\br{{\bm r}}
\def\bp{{\bm p}}
\def\bpi{{\bm \pi}}
\def\bM{{\bm M}}
\def\bs{{\bm s}}
\def\bS{{\bm S}}
\def\bB{{\bm B}}
\def\bE{{\bm E}}
\def\bA{{\bm A}}
\def\bj{{\bm j}}
\def\bF{{\bm F}}
\def\ux{{\bm e}_x}
\def\uy{{\bm e}_y}
\def\uz{{\bm e}_z}
\def\fe{{\mathcal F}}
\def\bI{{\bm I}}

\def\id{{\rm d}}
\def\bOm{{\bm \Omega}}
\def\bH{{\bm H}}

\def\br{{\bm r}}
\def\bv{{\bm v}}

\def\half{\frac{1}{2}}
\def\args{(\bm, t)}

\def\rdw{r_{\rm{dw}}}
\def\phidw{\varphi_{\rm{dw}}}
\def\ldw{\lambda_{\rm{dw}}}
\def\thetadw{\theta_{\rm{dw}}}

\def\sech{{\rm sech}}
\textit{Introduction} ---
Spintronic devices make use of the spin degree of freedom to process and store information. Hence, the generation and detection of non-equilibrium spin accumulation and spin currents is of paramount importance. In particular, all-electric injection and control of spin currents at room temperature and without high magnetic fields is crucial for viable integration with and as extension of current technology \cite{Jansen2012}. To obtain spin currents, a large variety of physical mechanisms and geometries are investigated \cite{Wolf2001,Takahashi2008,*Zutic2004}.  One class of approaches relies on parametric pumping \cite{Brouwer1998}. In these pumping approaches a periodic (AC) excitation is transformed into a DC spin current. Examples are circularly polarized optical photons \cite{Lampel1968,Ganichev2001,Stevens2003,Zutic2004}, magnons \cite{Sandweg2011,Ando2011,Kurebayashi2011}, acoustic waves \cite{Weiler2012} and single-domain ferromagnetic resonance \cite{Tserkovnyak2002,Tserkovnyak2002b,Mizukami2002}.

In contrast to the pumping approaches described above, a spin accumulation can also be obtained via a static bias. A current through a ferromagnetic-nonmagnetic-metal (FM-NM) junction causes spin injection into the nonmagnetic layer \cite{JohnsonSilsbee1987,*JohnsonSilsbee1988,Jedema2003}. Room temperature injection in silicon was demonstrated in Ref. \cite{Dash2009}. Spin-orbit coupling also opens the possibility to create spin currents using electric fields only. For example, the spin Hall effect generates a spin current transverse to a charge current \cite{Dyakonov1971,*Dyakonov1971b,*Hirsch1999,Sih2006, Murakami2004, Sinova2004}.

The interplay between charge, spin and temperature, studied in the young field of spin caloritronics, yields novel ways to inject pure spin currents using temperature gradients. Thermal transport leads to the generation of spin accumulation via the spin Seebeck effect \cite{Uchida2008}, spin dependent Seebeck effect \cite{Slachter2010} and  spin Seebeck tunneling \cite{LeBreton2011}.
\begin{figure}[t]
\begin{center}
\includegraphics[width=1.00\linewidth]{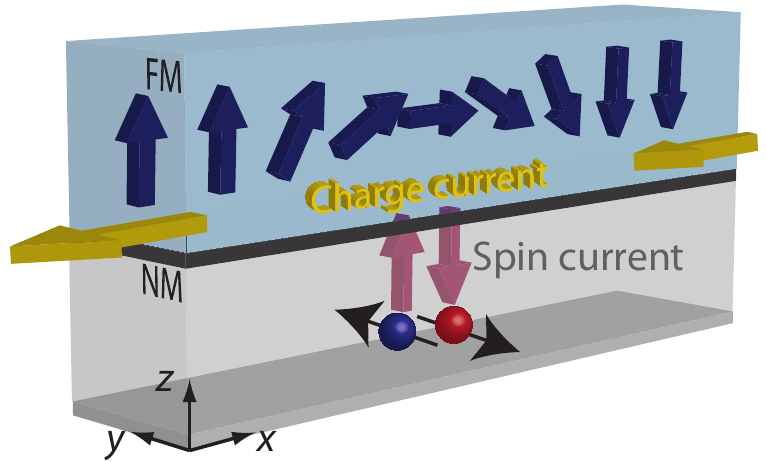}
\caption{(Color online) A charge current along the interface between a ferromagnet containing a domain wall and a normal metal leads to the injection of a transverse pure spin current at the interface. The direction of the spin component of the injected spin current is in the direction of the helicity $\bma\times(\bj_c\cdot\nabla)\bma$ of the domain wall along the current direction, leading to a spin-accumulation in the normal metal.} \label{fig:SpinInjection}
\end{center}
\end{figure}

In this Letter we propose a mechanism for the injection of a transverse spin current from a FM-NM interface by a charge current, as illustrated in Fig. \ref{fig:SpinInjection}. Consider a (pinned) domain wall in the FM layer such that the direction of magnetization $\bma(\bx)$ depends on the coordinate along the wire, and such that an electric current flows along the wire. We denote the charge current density on the interface by  $\bj_{c}$. It is the magnitude of this charge current density that governs the magnitude of the injected spin current. As we discuss in detail below, the change in orientation of the magnetization due to the presence of the domain wall will cause a transverse spin current $\bj^{\rm{in}}_{s}$ into the NM given by
\begin{equation}\label{eq:PhenoSpinIn}	
\bj^{{\rm in}}_{s}( \bx ) = \frac{\hbar g}{4\pi G_0} \bma( \bx )\times( \bv_s \cdot \nabla) \bma( \bx ),
\end{equation}
where $\bj^{{\rm in}}_{s}(\bx)$ is the injected spin-current density flowing perpendicular to the interface (and hence transverse to the charge current direction) with spin polarization in the direction of the helicity $\bma \times (\bj_c \cdot \nabla) \bma$  of the magnetization texture along the current direction (in the y-direction for the situation in Fig.\ref{fig:SpinInjection}). Here, $\bma(\bx)$ is the unit-vector magnetization direction of the FM-layer, $\bv_s=-g_L\mu_B P \bj_c/2 M_s |e|$ is the spin velocity, $P$ is the spin polarization in the ferromagnet, $g_L$ the Land\'e g-factor, $\mu_B$ is the Bohr magneton, $M_s$ is the saturation magnetization, and $-|e|$ the charge of an electron. Furthermore, the parameter $g=\eta g^{\uparrow \downarrow}$ is governed by properties of the FM-NM interface, where $g^{\uparrow \downarrow}$ is the real part of the mixing conductance in units ${\rm \Omega}^{-1}{\rm m}^{-2}$, and $\eta =\mathcal O \left( 1 \right)$ is  a dimensionless parameter and $G_0=2e^2/h$ is the quantum of conductance. Eq.~(\ref{eq:PhenoSpinIn}) is the main equation describing the effect we propose in this paper.

In the next section derive an expression for the coefficient $g$ in terms of spin-dependent scattering properties of the FM-NM metal interface, which can in principle be evaluated for realistic interfaces. This derivation is somewhat technical and some readers may wish to skip to the end of this section where we also give an estimate for $g$ using a physical interpretation of the transverse spin current in terms of spin pumping. 


\textit{Scattering Theory} ---
We consider a two-dimensional tight-binding model for the interface between ferromagnet and normal metal. The ferromagnetism is described with localized magnetic moments exchange coupled to the electrons, and connected on every site ${\bf r}=\{i_x,i_y\}$ to a metallic reservoir (i.e., the normal metal) with chemical potentials $\mu_{{\bf r}}$. The system is described by the Hamiltonian ${\cal H}={\cal H}_S+{\cal H}_L+{\cal H}_C$ representing electronic system in the FM, NM-leads and contacts between them. The first term is specified in terms of second-quantized operators $\hat{c}_{{\bf r},\sigma} (\hat{c}^{\dagger}_{{\bf r},\sigma})$ that annihilate (create) an electron with spin $\sigma$ at site ${\bf r}$
\begin{align}
{\cal H}_{S}=-J_{S}\sum_{\langle {\bf r},{\bf r}'\rangle;\sigma}\hat{c}^{\dagger}_{{\bf r},\sigma}\hat{c}_{{\bf r}',\sigma}-\Delta\sum_{{\bf r};\sigma,\sigma'}\hat{c}^{\dagger}_{{\bf r},\sigma}\bma_{{\bf r}}\cdot{\bm \tau}_{\sigma,\sigma'}\hat{c}_{{\bf r},\sigma'},
\end{align}
which describes nearest-neighbor (indicated by the brackets $\langle \cdot, \cdot \rangle$) hopping with amplitude $J_{S}$ and coupling to the local magnetic moments $\bma_{{\bf r}}$, where $\Delta$ is the exchange energy and $\bm{\tau}$ is the Pauli spin-matrix vector. The metallic contact is described by a set of one-dimensional leads at chemical potential $\mu_{\bf r}$ and modeled by the Hamiltonian ${\cal H}_L=\sum_{{\bf r}}{\cal H}_{{\bf r}}$ where the Hamiltonian for lead-${\bf r}$ is given by
${\cal H}_{{\bf r}}=-J_{L_{\bf r}}\sum_{\langle j',j''\rangle;\sigma}\left[\hat{d}^{L_{{\bf r}}}_{j',\sigma}\right]^{\dagger}\hat{d}^{L_{{\bf r}}}_{j'',\sigma}$, where the hopping amplitude is $J_{L_{{\bf r}}}$ and $\hat{d}^{L_{{\bf r}}}_{j,\sigma}$ and $\left[\hat{d}^{L_{{\bf r}}}_{j,\sigma}\right]^{\dagger}$ are  the fermionic operators in the ${{\bf r}}^{\rm th}$ lead. Finally the contact between the ferromagnetic system and its leads is described by ${\cal H}_C=\sum_{{\bf r}}{\cal H}^{{\bf r}}_{C}$ with
\begin{align}
{\cal H}^{{\bf r}}_{C}=-J^{{\bf r}}_C\sum_{\sigma}\left[\hat{c}^{\dagger}_{{\bf r},\sigma}\hat{d}^{L_{{\bf r}}}_{\partial L_{{\bf r}},\sigma}+\left[\hat{d}^{L_{{\bf r}}}_{\partial L_{{\bf r}},\sigma}\right]^{\dagger}\hat{c}_{{\bf r},\sigma}\right],
\end{align}
where $\partial L_{{\bf r}}$ denotes the last site of the lead and $J^{{{\bf r}}}_{C}$ is the amplitude of tunneling among both subsystems. An electric current flows through the FM by a difference in chemical potentials $\mu_{L,R}$ that connect the left and right sides of the FM.

The spin current flowing from the system to the leads is determined as the rate of change of spin density in the ${\bf r}$-th site, i.e. $\frac{d{\bf s}^{{\bf r}}}{dt}=\frac{\hbar}{2a^2}\frac{d}{dt}\sum_{\sigma,\sigma'}\langle\hat{c}^{\dagger}_{{\bf r},\sigma}{\bm \tau}_{\sigma,\sigma'}\hat{c}_{{\bf r},\sigma'}\rangle$, where $a$ is the distance between sites. The general expression has been derived before \cite{Duine2007} and is given by
\begin{eqnarray}\label{eq:spincurrentgeneralexpression}
\frac{d{\bf s}^{{\bf r}}}{dt}&=&\frac{i\hbar}{2a^2}\int\frac{d\epsilon}{2\pi}\text{Tr}\left[N\left(\epsilon-\mu^{{\bf r}}\right)\Gamma^{{\bf r}}_{{\bf r},{\bf r}}(\epsilon)\left({\bm \tau}G^{(+)}_{{\bf r},{\bf r}}(\epsilon)-G^{(-)}_{{\bf r},{\bf r}}(\epsilon){\bm \tau}\right)\right.\nonumber\\
&-& \left.\sum_{{\bf r}'}N\left(\epsilon-\mu^{{\bf r}'}\right)\left({\bm\tau}\Sigma^{{\bf r},(+)}_{{\bf r},{\bf r}}(\epsilon)-\Sigma^{{\bf r},(-)}_{{\bf r},{\bf r}}(\epsilon){\bm\tau}\right)A^{{\bf r}'}_{{\bf r},{\bf r}}(\epsilon)\right],
\end{eqnarray}
where $N (\epsilon) = \left[e^{\epsilon/k_BT}+1\right]^{-1}$ is the Fermi-Dirac distribution function with $k_B T$ the thermal energy, $A^{{\bf r}}(\epsilon)=G^{(+)}(\epsilon)\hbar\Gamma^{{\bf r}}(\epsilon)G^{(-)}(\epsilon)$ is the spectral-weight contribution due to the lead at site ${\bf r}$ and the rate $\Gamma^{{\bf r}}(\epsilon)=i \left[\Sigma^{{\bf r},(+)}(\epsilon)-\Sigma^{{\bf r},(-)}(\epsilon)\right]$. Since we are assuming non-magnetic leads, the self-energy $\Sigma^{{\bf r},(+)}(\epsilon)$ will be proportional to the identity in spin space. Its only non-zero matrix elements are $\hbar\Sigma^{{\bf r},(+)}_{{\bf r},\sigma;{\bf r},\sigma'}=-(J^{{\bf r}}_C)^2 e^{ik^{{\bf r}}(\epsilon)a}\delta_{\sigma,\sigma'}/J_{L_{{\bf r}}}$, with $k^{{\bf r}}(\epsilon)a=\arccos\left[-\epsilon/2J_{L_{{\bf r}}}\right]$. To carry out the explicit evaluation of Eq.~(\ref{eq:spincurrentgeneralexpression}) in terms of the magnetization orientation $\bma_{\bf r}$ it is convenient to decompose the Green's functions into spin-independent {\it singlet} and spin-dependent {\it triplet} parts, namely
\begin{align}
G^{(\pm)}_{{\bf r},\sigma;{\bf r}',\sigma'}(\epsilon)=G^{(s)(\pm)}_{{\bf r},{\bf r}'}(\epsilon)\delta_{\sigma,\sigma'}+G^{(t)(\pm)}_{{\bf r},{\bf r}'}(\epsilon)\bma_{{\bf r}}\cdot{\bm \tau}_{\sigma,\sigma'}.
\end{align}
Taking the trace over the spin indices in Eq.~(\ref{eq:spincurrentgeneralexpression}) we distinguish two contributions to the spin current, one component parallel to the magnetization vector and other transverse, denoted by ${\bj}^{\parallel}_{s,{\bf r}}$ and ${\bj}^{\perp}_{s,{\bf r}}$ respectively. The transverse spin current density induced by the magnetic texture to lead ${\bf r}$ is given by
\begin{equation}\label{eq:transversespincurrent}
{\bj}^{\perp}_{s,{\bf r}}=\frac{1}{a^2}\int\frac{d\epsilon}{(2\pi)}\sum_{{\bf r}'}N_{\rm FD}\left(\epsilon-\mu^{{\bf r}'}\right)\text{t}^{(t)}_{{\bf r}{\bf r}'} (\epsilon)  \left(\bma_{{\bf r}}\times\bma_{{\bf r}'}\right),
\end{equation}
with the transmission probability  $\text{t}^{(t)}_{{\bf r}{\bf r}'} (\epsilon)=\hbar\Gamma^{{\bf r}}_{{\bf r},{\bf r}} (\epsilon) G^{(t)(+)}_{{\bf r},{\bf r}'} (\epsilon) \hbar\Gamma^{{\bf r}'}_{{\bf r}',{\bf r}'} (\epsilon) G^{(t)(-)}_{{\bf r}',{\bf r}} (\epsilon)$ for the spin-polarized part of the current flowing from the lead at site ${\bf r}$ to the lead at site ${\bf r}'$ through the FM. We now consider a zero net current (but nonzero spin current) flow into the leads, except for the left and right leads that have chemical potentials $\mu_{L}=\epsilon_{F}+|e|V$ and $\mu_{R}=\epsilon_{F}$, respectively, with $\epsilon_F$ the Fermi energy. At low temperatures and assuming the length scale of magnetization-orientation variation much greater than the inverse Fermi wavelength we see that in the continuum limit the only contributions to Eq.~(\ref{eq:transversespincurrent}) are from neighboring leads for which ${\bm m}_{\bf r}\times {\bm m}_{{\bf r}'}\rightarrow {\bm m}({\bf x})\times  a\partial_x{\bm m}({{\bf x}})$. Keeping these contributions, we find that the transverse spin-current density to lowest order in magnetization gradients satisfies Eq.~(\ref{eq:PhenoSpinIn}) with $g = a (16 M_s/g_L \mu_B P) \times (G_0 \text{t}^t (\epsilon_F)/\text{t} (\epsilon_F))$, where $\text{t} (\epsilon_F)$  is the total (i.e., summed for both spin channels) transmission probability. Both this transmission probability, and the spin-dependent transmission probability are taken between leads at neighboring site and are taken at the Fermi energy in the low-temperature limit. Also note that these transmission probabilities are evaluated for the homogeneous ferromagnetic state, as Eq.(\ref{eq:transversespincurrent}) is already first order in magnetization gradient.

We obtain an estimate for the coefficient $g$ by connecting the transverse spin current [c.f. Eq. \eqref{eq:PhenoSpinIn}] to spin pumping by a precessing magnetization. To make this connection we consider a helical magnetization $\bma_{\rm hel}(q x + \omega t)$ precessing with frequency $\omega$. We assume the wavelength of the helix to be much larger than the spin diffusion length in the ferromagnet, $q\lambda_{\rm sd}\ll 1$, such that we can use the spin-pumping expression for a single-domain ferromagnet\cite{Tserkovnyak2002,Tserkovnyak2002b}
\begin{equation}\label{eq:Pump}
\bj^{\rm pump}_{s} = \frac{\hbar g^{\uparrow \downarrow}}{4 \pi G_0} \bma\times\frac{\partial \bma}{\partial t},
\end{equation}
for the spin current $\bj^{\rm pump}_s$ pumped through the FM-NM interface by a time-dependent magnetization. Now we make a Galilean transformation to the frame moving with velocity $\omega/q$. In this frame we have a static magnetization and a spin velocity of $-\omega/q$ and obtain Eq. (\ref{eq:PhenoSpinIn}) with $g/g^{\uparrow\downarrow}=1$ by demanding that the injected spin current does not change. Note that Galilean invariance is in general broken due to disorder leading to a $g/g^{\uparrow\downarrow}\neq 1$ but of order unity. From this argument we can also make an estimate for the magnitude of the transverse spin current injected by a domain wall of width $\lambda_{\rm dw}$. For narrow domain walls with $\lambda_{\rm dw}$ of the order of a few nm and currents for which $v_s$ is of the order of $100$ m$/$s (that can be obtained in experiments), the injected spin current density would be equivalent to pumping with a ferromagnetic resonance frequency of $v_s/\lambda_{\rm dw}\approx 100$ GHz.

\textit{Inverse Spin Hall Effect} ---
%
We now consider how the transverse spin current can be probed by the inverse spin Hall effect (ISHE). We consider a pinned domain wall with a width $\lambda_{{\rm dw}}$ in a perpendicular magnetized FM, as shown in Fig. \ref{fig:SpinInjection}, where $\varphi_{\rm dw}$ is the angle of the magnetic moments in the wall with the $x$-axis. We model the magnetization by $\bma_{\rm dw}(x)=\left(\cos\varphi_{\rm dw}\sin\theta,\sin\varphi_{\rm dw}\sin\theta,\cos\theta \right)^T,$
where $\theta = 2\arctan \exp\left(Q\frac{x-r_{\rm dw}}{\lambda_{\rm dw}}\right)$, and $Q$ is the charge of the domain wall. The injected spin-current will result in a nonzero spin accumulation ${\bm \mu}_s(x,z)$ which can be found by solving the spin-diffusion equation in the NM \cite{Tserkovnyak2002b}, which in the static limit $\tau_{\rm sf} \partial{\bm \mu}_s/\partial t\ll 1$, is given by
\begin{equation}\label{eq:SpinDiffusion}
\nabla^2{\bm \mu}_s = \frac{{\bm \mu}_s}{\lambda_{\rm sd}^2},
\end{equation}
where $\lambda_{\rm sd}\equiv \sqrt{D_s \tau_{{\rm sf}}}$ is the spin-diffusion length in the NM, and $D_s$ and $\tau_{{\rm sf}}$ are its spin-diffusion constant and spin-flip time, respectively. The boundary conditions for Eq. (\ref{eq:SpinDiffusion}) enforce continuity for the spin current and are given by
\begin{eqnarray}
\left.\partial_z{\bm \mu}_s(x,z)\right|_{z=0} &=& -\frac{G_0}{\sigma}\bj_{s,z}^{\rm net}(x);\\
\left.\partial_z{\bm \mu}_s(x,z)\right|_{z=d_N} &=&0;\\
\left.\partial_x{\bm \mu}_s(x,z)\right|_{x=\pm L_N} &=&0,
\end{eqnarray}
where $L_N$, $d_N$ and $\sigma$ are the length, thickness and conductivity of the NM respectively, $\bj_{s,z}^{\rm net}(x)$ is the net spin-current into the NM. The solution for the spin accumulation yields $ {\bm \mu}_s(x,z) = \int dx' K(x-x',z) \bj_{s,z}^{\rm net}(x') $ where the fourier transform of the kernel $K(x,z)$ is given by
\begin{equation}\label{eq:SpinAcc}
\tilde{K}(k_x,z)=-\frac{G_0\lambda_{sd}}{\sigma}\frac{\cosh\left(\frac{z+d_N}{\lambda_{sd}}\sqrt{k_x^2\lambda_{sd}^2+1}\right)}{\sqrt{k_x^2\lambda_{sd}^2+1}\sinh\left(\frac{d_N}{\lambda_{sd}}\sqrt{k_x^2\lambda_{sd}^2+1}\right)}.
\end{equation}
The total spin current is the sum of the spin-injection and spin-pumping contributions, as given in Eqs. (\ref{eq:PhenoSpinIn}) and (\ref{eq:Pump}), and a spin-current $j_s^{\rm back}$ in the opposite direction due to the induced spin-accumulation on the NM side of the interface. Thus  $\bj^{{\rm net}}_{s,z} = \bj^{{\rm in}}_{s,z}+\bj^{\rm pump}_{s,z}+\bj^{{\rm back}}_{s,z}$, where the back-flow of spins trough the interface is given by\cite{Brataas2001}
\begin{equation}\label{eq:Back}
\bj_{s,z}^{{\rm back}} \approx \frac{g^{\uparrow \downarrow}}{4\pi G_0} {\bm \mu}_s(x,z=0),
\end{equation}
where we neglected the imaginary part of the mixing conductance. This equation is justified for realistic interfaces\cite{Xia2002}. Using Eq. (\ref{eq:SpinAcc}) in Eq. (\ref{eq:Back}) we get an expression for the total spin-current in terms of the spin-injection and pumping contributions only. Note that here the pumping contribution is zero as we consider a static pinned domain wall. We come back to this contribution when we consider the interfacial enhancement of spin transfer torques. 

The ISHE \cite{Saitoh2006,Kimura2007} gives a voltage signal from a spin current via $j_{c,i}^{\rm ISHE} =\frac{2e}{\hbar} \theta_{\rm SH} \epsilon^{ij\alpha}j_{s,j}^{\alpha}$, with the spin current given by  $j_{s,j}^{\alpha}=\frac{\sigma}{G_0} \partial \mu_s^{\alpha} /\partial x^j $ and $\theta_{\rm SH}$ the spin-Hall angle. We average the voltage generated via the ISHE over the thickness $d_N$ of the NM-layer. The voltage difference due to the ISHE in the $x$-direction is then
\[
\frac{\Delta V_x^{\rm ISHE} }{\eta}=  \frac{\hbar v_s}{2 d_N |e|} \theta_{\rm SH}\frac{(\cosh\frac{d_N}{\lambda_{\rm sd}}-1)Q \cos\varphi_{\rm dw} }{\frac{4\pi \sigma}{g^{\uparrow \downarrow}\lambda_{\rm sd}} \sinh \frac{d_N}{\lambda_{\rm sd}}-\cosh\frac{d_N}{\lambda_{\rm sd}}},
\]
which holds for the domain wall far from the edges of the NM-layer. Note the thickness of the Pt-layer influences the signal considerably\cite{Castel2012}, but that the result does not depend on the width of the domain wall because the ISHE induced voltage is in the x-direction. For
a Co-Pt interface we take $g^{\uparrow\downarrow} \approx4\cdot 10^{15} {\rm \Omega}^{-1}{\rm m}^{-2}$\cite{Czeschka2011}, $\sigma_{\rm Pt} = 9.5\cdot 10^6 {\rm \Omega}^{-1}{\rm m}^{-1}$, $\lambda_{\rm sd}=1.5 {\rm nm}$, $\lambda_{\rm dw}\approx 5{\rm nm}$ and a platinum thickness of $d_N = 3{\rm nm}$, $\theta_{\rm SH}=.05$ and obtain $\Delta V_x^{\rm ISHE}/\eta \approx 40 \rm{nV}$. For a current density in the cobalt of $j_c \approx 10^{12} {\rm A m}^{-2}$.

\begin{figure}[t]
\begin{center}.
\includegraphics[width=1.00\linewidth]{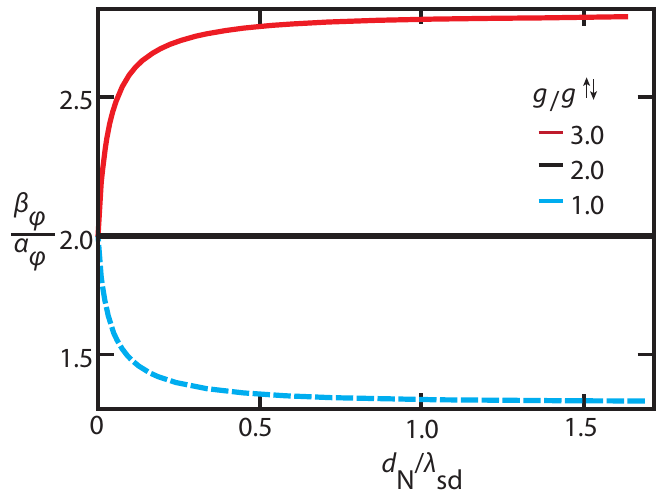}
\caption{(Color online)  The ratio $\beta_{\varphi}/\alpha_{\varphi}$ is shown as a function of the thickness $d_{\rm N}$ of the normal metal layers in units of its spin diffusion length $\lambda_{\rm sd}=250{\rm nm}$, for a Cu-Py-Cu multilayer. The head-to-head domain wall in the Py has a width of 100 nm. For $g/g^{\uparrow\downarrow}=2$, where $g^{\uparrow\downarrow}=1.6\cdot10^{15}{\rm \Omega}^{-1}{\rm m}^{-2}$ is the mixing conductance of a Py-Cu interface the ratio, the ratio  $\beta_\varphi/\alpha_\varphi = 2$ coincides with the bulk ratio with $\alpha_0=0.006$.} 
\label{fig:betaoveralpha}
\end{center}
\end{figure}

\textit{Interfacial enhancement of spin transfer} ---
Another way to observe the injection of the spin current is the effect it has on the motion of the domain wall. This is important for the understanding of domain-wall dynamics in layered magnetic materials, that are the subject of ongoing research \cite{Hayashi2007}. The total spin current ejected from the FM-layer reduces the angular momentum of the FM, inducing a torque which modifies the domain-wall dynamics.  The Landau-Lifshitz-Gilbert equation in the presence of this additional torque is given by
\begin{eqnarray}\label{eq:LLG}
\frac{\partial \bma}{\partial t} +v_s\frac{\partial \bma}{\partial x} &=& \alpha_0\bma\times\frac{\partial \bma}{\partial t}-\frac{1}{\hbar}\bma\times\frac{\partial E_{\rm MM}}{\partial \bma} \nonumber\\
 &+&\beta_0 v_s\bma\times\frac{\partial \bma}{\partial x}+\frac{\gamma}{M_s d_F}\bj_{s,z}^{\rm net},
\end{eqnarray}
where $\alpha_0$ and $\beta_0$ are the bulk Gilbert damping and non-adiabaticity parameter respectively. The FM has thickness $d_F$ and hard-axis anisotropy $K_{\perp}$ which we take to be along the y-axis and is included in the energy functional $E_{\rm MM}$ that also contains the exchange and easy axis anisotropy that set the domain-wall width. From the LLG equation [Eq. (\ref{eq:LLG})] we obtain the equations of motion for the collective coordinates of the domain wall, which are given by
\begin{eqnarray}
\dot{\varphi}_{\rm dw}+\alpha_{\varphi} \frac{\dot{r}_{\rm dw}}{\lambda_{\rm dw}}&=&\beta_{\varphi}\frac{v_s}{\lambda_{\rm dw}};\\
\frac{\dot{r}_{\rm dw}}{\lambda_{dw}}-\alpha_{r} \dot{\phi}_{\rm dw}&=& \frac{K_{\perp}}{2\hbar}\sin2\varphi_{\rm dw}+\frac{v_s}{\lambda_{\rm dw}},
\end{eqnarray}
where $\alpha_{r,\varphi}$ and $\beta_{\varphi}$ are given by
\begin{eqnarray}
\alpha_{r,\varphi}&=&\alpha_0+\frac{\gamma\hbar g^{\uparrow \downarrow}}{4\pi M_s d_F}\mathcal{I}_{r,\varphi}  \quad;\quad \beta_{\varphi}=\beta_0 +\eta\frac{\gamma \hbar g^{\uparrow \downarrow}}{4\pi M_s d_F}\mathcal{I}_{\varphi};\nonumber\\
\mathcal{I}_{\xi} &=&\frac{4\pi G_0}{\hbar g^{\uparrow \downarrow}}\iint dx dx' \, \Pi(x-x') \bj_{s,z}^{{\rm in}+{\rm pump}}(x)\cdot\frac{\delta \bma_{\rm dw}}{\delta \xi}(x'); \nonumber\\
\end{eqnarray}
with $\xi =\{r,\varphi \}$ and where $\Pi(x-x')$ is given by
\begin{eqnarray}
\Pi(x-x')&=& \int_{-\infty}^{\infty} \frac{dk_x}{2\pi} \left[1-\frac{g^{\uparrow \downarrow}}{4\pi G_0}\tilde{K}(k_x,0) \right]^{-1}e^{ i k_x (x-x')}.\nonumber
\end{eqnarray}
Note that the integrals $\mathcal{I}_{\varphi,r}$ are functions of the dimensionless parameters $\lambda_{\rm dw}/\lambda_{\rm sd}$, $d_N/\lambda_{\rm sd}$ and $g^{\uparrow \downarrow}\lambda_{{\rm sd}}/4\pi\sigma$. The average velocity of the current-driven domain wall is given by
\[
\left\langle \dot{r}_{\rm dw} \right\rangle=\frac{ \beta_{\varphi}}{\alpha_{\varphi}}v_s+\frac{{\rm sign}\left[1-\frac{\beta_{\varphi}}{\alpha_{\varphi}}\right]}{1+\alpha_r\alpha_{\varphi}}{\rm Re}\left[\sqrt{(1-\frac{\beta_{\varphi}}{\alpha_{\varphi}})^2v_s^2-v_c^2}\right],
\]
where the critical velocity is given by $v_c=K_{\perp}\lambda_{\rm dw} /2\hbar$. The ratio $\beta_{\varphi}/\alpha_{\varphi}$ determines the qualitative behavior of the domain-wall velocity as a function of current. In Fig. \ref{fig:betaoveralpha} this ratio is shown as a function of the thickness of the normal metal layers in the multilayer. Note that in the limit of a large ratio $\lambda_{\rm dw}/\lambda_{\rm sd}$ our result for alpha coincides with interfacial enhancement of Gilbert damping in single-domain magnets obtained by Tserkovnyak \textit{et al.} \cite{Tserkovnyak2002,Tserkovnyak2002b}.
%

\textit{Discussion and Conclusions} ---
In this Letter we described a mechanism for transverse spin injection across a FM-NM interface that is induced by a magnetization gradient along a current, and focused on the example of a pinned domain wall. We note here that if the domain wall is not pinned and moves with velocity $v_{\rm dw}$, the injected spin current is altered which is described by the replacement $v_s \rightarrow v_s - v_{\rm dw}$ in Eq. (\ref{eq:PhenoSpinIn}). The position of the injection is dependent on the position of the domain wall which could lead to controllable local spin injection. Moreover, due to the spin accumulation induced in the NM layer(s) the measured $\alpha$ and $\beta$ parameters for domain-wall motion for multilayer systems will be different than bulk values. We expect that this effect plays a role in all thin-film measurements of domain-wall dynamics. In fact, large values of beta are typically reported in such systems \cite{Miron2009}, pointing to the possibility of interfacial enhancement.

Other spin-injection mechanisms, like the spin-dependent Seebeck effect\cite{Slachter2010} or diffusive spin injection\cite{Jedema2003}, typically induce a spin current in the NM-layer with the spin direction parallel to the magnetization in the ferromagnet. Therefore the spin-injection we discussed in this Letter, which induces a spin current polarized in the direction of the helicity, is distinguishable.

We also note that in principle there is a spin current with spin polarization in the direction of $({\bm v}_s\cdot\nabla)\bma$ that we have ignored as it oscillates and averages out to small values when integrated over position, and is determined approximately  by the imaginary part of the mixing conductance which is small for realistic interfaces.

The domain-wall induced spin injection could be used in a memory device consisting of a $\rm{FM}_1\rm{-NM-FM}_2$ trilayer where $\rm{FM}_1$ is a ferromagnet with a domain wall adjacent to $\rm{FM}_2$, which is a monodomain. For the geometry as shown in Fig. \ref{fig:SpinInjection} the injected spin-current has the same symmetry as the SHE. We can estimate the magnitude of the effective spin Hall angle for the spin-injection to be $\theta=|2 e j_s/\hbar j_c|\approx 0.008$ for a Co-Pt interface, where the spin current was estimated by replacing the gradient in Eq. (\ref{eq:PhenoSpinIn}) by $1/\lambda_{\rm dw}$. The presence of this spin current in addition to the ISHE spin current could be used to control the switching of the single domain ${\rm FM}_2$ by the presence of a domain wall in ${\rm FM}_1$.

The reverse process of spin-injection occurs as well. A spin current flowing with a transverse spin direction  into a ferromagnet at the position of the domain wall will lead to a voltage difference over the ferromagnetic strip. In this way domain walls could be used as a local moveable sensors of spin current.
In future work we plan to investigate transverse pure spin currents associated with heat currentsin the same geometry as discussed here.

\acknowledgements
It is a pleasure to thank Gerrit Bauer, Gavin Burnell, Bryan Hickey, Mathias Kl\"aui, Chris Marrows, Tom Moore, and Maxim Tsoi for discussions. This work was supported by the Stichting voor Fundamenteel Onderzoek der Materie (FOM), the Netherlands Organization for Scientifc Research (NWO), and by the European Research Council (ERC).
\bibliography{Domain-wall_controlled_transverse_spin_injection_v3.bbl}
\end{document}